\documentclass[a4paper]{article}
\usepackage{INTERSPEECH2020}
\usepackage[export]{adjustbox}
\usepackage{subfigure}
\usepackage{hyperref}
\usepackage[shortlabels]{enumitem}
\usepackage{CJKutf8}
\usepackage{graphicx}

\usepackage{algorithm}  
\usepackage{algpseudocode}  
\usepackage{amsmath}  

\hypersetup{
    colorlinks=false,
    linkcolor=blue,
    filecolor=blue,      
    urlcolor=blue,
    citecolor=cyan,
}

\title{Controllable  Context-aware Conversational Speech Synthesis}
\name{Jian Cong$^{1}$\thanks{Part of this work performed when Jian Cong was interning at Tencent AI Lab. Lei Xie is the corresponding author.}, Shan Yang$^2$, Na Hu$^2$, Guangzhi Li$^2$, Lei Xie$^1$, Dan Su$^2$}
\address{
  $^1$Audio, Speech and Language Processing Group (ASLP@NPU), School of Computer Science, Northwestern Polytechnical University, Xi'an, China\\
  $^2$ Tencent AI Lab, China}
\email{npujcong@mail.nwpu.edu.cn, \{shaanyang,ninahu,guangzhilei,dansu\}@tencent.com \\ lxie@nwpu.edu.cn}

\begin{document}
\begin{CJK*}{UTF8}{gbsn}

\maketitle
\begin{abstract}

In spoken conversations, spontaneous behaviors like filled pause and prolongations always happen. Conversational partner tends to align features of their speech with their interlocutor which is known as entrainment. To produce human-like conversations, we propose a unified controllable spontaneous conversational speech synthesis framework to model the above two phenomena. Specifically, we use explicit labels to represent two typical spontaneous behaviors \textit{filled-pause} and \textit{prolongation} in the acoustic model and develop a neural network based predictor to predict the occurrences of the two behaviors from text. We subsequently develop an algorithm based on the predictor to control the occurrence frequency of the behaviors, making the synthesized speech vary from less disfluent to more disfluent. To model the speech entrainment at acoustic level, we utilize a context acoustic encoder to extract a global style embedding from the previous speech conditioning on the synthesizing of current speech. Furthermore, since the current and previous utterances belong to the different speakers in a conversation, we add a domain adversarial training module to eliminate the speaker-related information in the acoustic encoder while maintaining the style-related information. Experiments show that our proposed approach can synthesize realistic conversations and control the occurrences of the spontaneous behaviors naturally.

\end{abstract}

\noindent\textbf{Index Terms}: Speech synthesis, Spontaneous speech, Conversational speech
\vspace{-0.4cm}
\section{Introduction}
With the wide use of deep neural networks (DNN), speech synthesis has been significantly advanced from conventional frame-wise linguistic-acoustic frameworks~\cite{black2007statistical,tokuda2000speech,ze2013statistical,ling2015deep,arik2017deep} to sequence-to-sequence (seq2seq) paradigms~\cite{wang2017tacotron,shen2018natural,li2019close}. 
Recently, there have been increasing interests in generating expressive speech with explicit or implicit style or emotion modeling~\cite{skerry2018towards, wang2018style,zhang2019learning,bian2019multi,lei2020fine}. However, there are only a few studies working on spontaneous conversational speech synthesis.
Although the current TTS has advanced in expressiveness, it's still a great challenge to imitate human conversations because it's hard to model the subtle spontaneous behaviors in conversations. Since humans always think while talking, the conversational speech often contains ``spontaneous behaviors`` like hesitation, interruption, breathing, or other disfluencies. These behaviors are also affected by the conversation context and the entrainment phenomenon from the two parties in conversations. This study works on spontaneous conversational speech synthesis by explicitly modeling the spontaneous behaviors and the conversation context, aiming at generating truly realistic conversations.


As for the spontaneous behaviors, some studies particularly focused on the insertion and synthesis of filler pauses (especially ``uh'', ``um'')~\cite{tomalin2015lattice, dall2014investigating, szekely2019casting}. As for the modeling of conversation history, researchers have found that using contextual statistical features can apparently improve naturalness in conversational speech synthesis~\cite{yamashita2020investigating,koriyama2011use}. Besides, the syntactic structure and chat history are also proved to be useful in recent seq2seq-based conversational speech synthesis~\cite{guo2020conversational}. Building appropriate corpus on conversational speech has also drawn much attention~\cite{guo2020conversational} while found data such as podcast is also a good resource for building a conversational TTS system~\cite{szekely2019casting}.

In this paper we propose a unified  controllable spontaneous conversational speech synthesis framework based on Tacotron2 to generate realistic two-party conversations. The proposed method specifically models 1) spontaneous behaviors and 2) conversation context between talkers -- the two essential aspects making the conversations more natural.
\begin{figure*}[t]
  \centering
  \setlength{\belowcaptionskip}{-20pt}
  \setlength{\abovecaptionskip}{10pt}
  \includegraphics[width=0.65\linewidth]{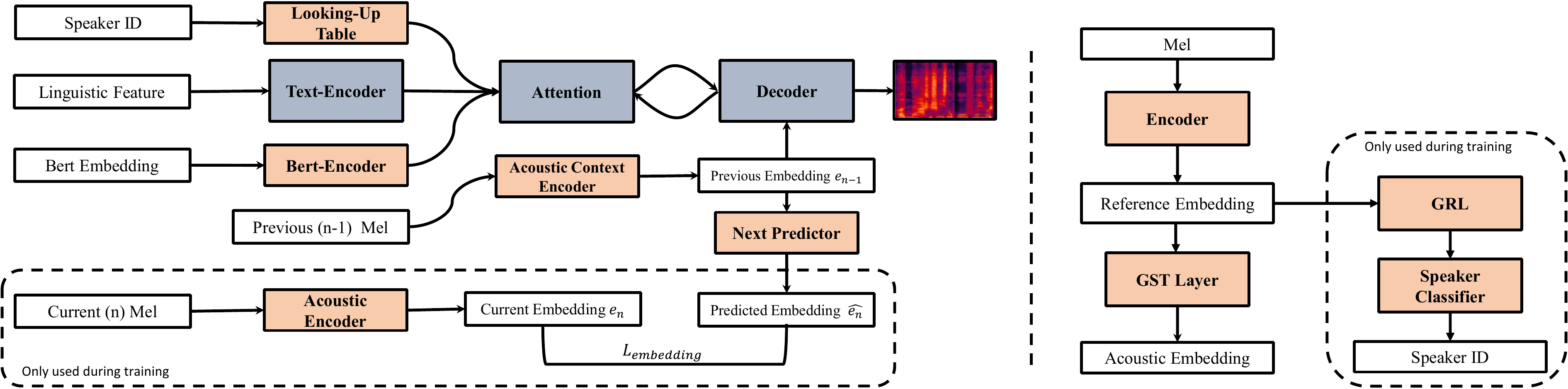}
  \caption{The left part is the context-aware based acoustic model. $L_{embedding}$ denotes the MSE loss between the current embedding $e_n$ and $\hat{e_n}$ predicted from previous embedding $e_{n-1}$. The right part shows the detailed architecture of the acoustic context encoder, which produces speaker-independent acoustic embeddings for chat history modeling. }
  \label{fig:am_architecture}
 \end{figure*}

We mainly focus on two typical acoustic phenomena caused by spontaneous behaviors, the \textit{prolongation} due to thinking or emphasis, and the \textit{filled-pause} due to repetition or stuttering. We treat each spontaneous behavior as an word attribute as these acoustic spontaneous behaviors may exist after any word in a sentence. In conversational speech synthesis, the spontaneous behavior tags are not readily available from texts, so we propose a new spontaneous behavior predictor based on a pre-trained BERT model. Different from all previous approaches, with the predictor, we further develop an algorithm to control the appearances of the spontaneous behaviors, making the speech vary from less disfluent to more disfluent. Experiments demonstrate that with the predicted spontaneous behaviors tags, the synthesized conversations are more natural compared with those using random tags. Moreover, with the predictor we can easily control the frequency of spontaneous behaviors occurring in the synthesized speech.

Previous work has suggested that modeling conversation history is beneficial to natural conversation synthesis~\cite{guo2020conversational}. Plenty of evidence has shown that the interlocutors in spoken conversations have been shown to entrain, or become similar to each other, in multiple dimensions~\cite{levitan2015entrainment}, including those at the acoustic/prosody level. Hence, in this work, we propose to inject an acoustic context encoder to model this important phenomenon. In detail, the acoustic encoder learns a fixed-length embedding from the previous utterance, which is utilized as a condition to the auto-regressive decoder of our acoustic model. Inspired by~\cite{oplustil2020using}, we also utilize the embedding from the previous utterance to predict the context embedding of the current utterance for additional supervision. However, since the previous sentence and the current sentence belong to the different speakers in a conversation, we add a domain adversarial training module to the acoustic encoder, which tends to eliminate the speaker information and only remain the style-related feature in the embedding space. Experiments show that our proposed methods significantly improve the performance of conversational speech synthesis.
\vspace{-0.3cm}
\section{Context-aware acoustic model}
Fig.~\ref{fig:am_architecture} illustrates the proposed context-aware acoustic model for spontaneous conversational speech synthesis. The architecture is based on Tacotron2~\cite{shen2018natural}.
Specifically, the model mainly consists of a CBHG-based text encoder~\cite{wang2017tacotron}, a BERT encoder, an acoustic context encoder and an attention-based auto-regressive decoder.

The text encoder firstly encodes phoneme-level features into linguistic representations $c={(c_1, c_2,... c_N)}$, while the decoder generates mel-spectrum $\hat{m} = {(m_1, m_2, m_M)}$ frame by frame with the linguistic representations $c$ in an auto-regressive manner and the MSE loss between the predicted $\hat{m}$ and the ground truth $m$ are minimized to optimize the whole model:
\begin{equation}
 \setlength{\abovedisplayskip}{3pt}
 \setlength{\belowdisplayskip}{3pt}
 \hat{m} = d(c | \Theta_{d}),~~~~L_{rcon} = ||m - \hat{m}||.
 \label{eq_decoder}
\end{equation}
where $d$ represents the decoder process, and $\Theta_d$ represents the model parameters of the decoder. 

Based on the above basic acoustic model, we inject a simple speaker look-up table to model both parties in the conversation. In order to directly control the generation of spontaneous behaviors, the explicit spontaneous behavior tags are used as input along with the phoneme-level linguistic features. As for the conversation history modeling, we build a speaker-independent acoustic context encoder to capture the chat histories. Finally, another auxiliary CBHG-based BERT encoder is adopted to help to model the semantic information from the pre-extracted BERT embeddings~\cite{devlin2018bert}.
\vspace{-0.2cm}
\subsection{Explicit representation of spontaneous behaviors}\label{sec:label}
We mainly focus on two common acoustic spontaneous behaviors--\textit{prolongation} and \textit{filled-pause}.
Specially, we define two linguistic features for these behaviors named \textit{prolongation} and \textit{filled-pause} in the text-side, where pronunciation prolongation may occur at the end of a character and \textit{filled-pause} may come right after the character. It is worth noticing that the \textit{filled-pause} here does not come from a normal rhythm change in fluent reading-style speech, and it is a spontaneous event may be inserted anywhere in an utterance. Together with the input phones, tones, and prosody labels, these phoneme-level linguistic features are utilized to predict the corresponding acoustic targets. We simply replicate them to form phoneme-level representations according to the character pronunciations.
\vspace{-0.2cm}
\subsection{Speaker-independent acoustic context encoder}
The current TTS systems are usually trained using sentence-level audio clips. Previous studies have shown that leveraging the prosody phenomena above the utterance level is beneficial to the naturalness of audiobook generation~\cite{oplustil2020using}. Specifically, simply considering the previous utterance by using an acoustic context encoder can lead to improved naturalness of synthesized audiobook. For spontaneous conversational speech synthesis, it is more reasonable to leverage the sequential nature of the two-party dialogue because of the entrainment phenomena -- the phenomenon of the speech of conversational partners becoming more similar to each other in multiple dimensions, including the important entrainment at acoustic/prosody level. Hence, we introduce the idea of acoustic context encoder to the synthesis of conversational speech. But differently, as conversations involve at least two speakers, we have to make the acoustic context encoder work in a speaker-independent manner which is realized by an adversarial learning module removing speaker identity information.

To describe the speech conversation process, we formulate a typical conversation process as:
\begin{equation}
 \setlength{\abovedisplayskip}{3pt}
 \setlength{\belowdisplayskip}{3pt}
 C = \{A_1, B_2, A_3, ... A_{(n-1)},B_n\}
 \label{Conv}
\end{equation}
where $C$ means a complete conversation with $n$ sentences from the two speakers A and B. The speech utterance $A_i$ indicates the $i$-th sentence in this conversation, which is from speaker $A$.  

In a conversation $C$, we assume the utterance $A_n$ is influenced by $B_{(n-1)}$, and  $B_{(n-1)}$ is influenced by $A_{(n-2)}$, etc. To make use of the information from the previous utterance when generating current sentence, we build a context acoustic encoder inspired by~\cite{oplustil2020using}, which encodes the previous context into a fixed-length acoustic embedding to inform the current sentence within the auto-regressive decoding. Assume the current training sample is $A_n$, and its previous sentence is $B_{(n-1)}$, the context embedding is obtained by:
\begin{equation}
 \setlength{\abovedisplayskip}{3pt}
 \setlength{\belowdisplayskip}{3pt}
 e_{n-1} = f_p(B_{(n-1)}|\Theta_{f_p})
 \label{eq_fp}
\end{equation}
where the $f_p$ is the acoustic context encoder, which produces context embedding $e_{n-1}$ from previous utterance.
Similar to the ``next task'' modeling in ~\cite{oplustil2020using}, we also build an acoustic encoder to extract the acoustic embedding from the current sentence $A_n$:
\begin{equation}
 \setlength{\abovedisplayskip}{3pt}
 \setlength{\belowdisplayskip}{3pt}
 e_{n} = f_c(A_n|\Theta_{f_c})
 \label{eq_fc}
\end{equation}
where $f_c$ is another acoustic encoder. With the acoustic embedding $e_{n-1}$ from the previous utterance and the embedding $e_n$ from the current utterance, we further adopt a ``next predictor'' to predict $e_n$ from $e_{n-1}$, which can model the relations between $e_{n-1}$ and $e_n$:
\begin{equation}
 \setlength{\abovedisplayskip}{3pt}
 \setlength{\belowdisplayskip}{3pt}
 \hat{e}_{n} = h(e_{n-1}|\Theta_h), ~~~ L_{embedding} = ||\hat{e}_{n}, e_{n}||_2
 \label{eq1}
\end{equation}
where $h$ is the ``next predictor", which contains two feed-forward layers. Note that the embedding $e_n$ is only used during training to form an additional loss.

In details, both the two acoustic encoders $f_p$ and $f_c$ follow the same architecture, as shown in the right part in Fig.\ref{fig:am_architecture}. It contains a reference encoder~\cite{skerry2018towards}, a speaker classification module with a gradient reverse layer (GRL), and a global style tokens (GST) layer~\cite{wang2018style}. The acoustic embedding is obtained from the output of the reference encoder. Here the GRL is used to disentangle the  speaker information in the embedding; otherwise, the acoustic encoder may easily learn the identity of the speaker in the conversation $C$ and this is not the case we want as we desire a speaker-independent acoustic embedding.

With the acoustic context modules above, the history embedding $e_{n-1}$ is fed to each state of in the auto-regressive decoder. Hence, the decoding process in Eq.~\eqref{eq_decoder} becomes
\begin{equation}
 \setlength{\abovedisplayskip}{3pt}
 \setlength{\belowdisplayskip}{3pt}
 \hat{m} = d(c, s, e_{n-1} | \Theta{d}), 
\end{equation} where $s$ represents speaker embedding. Combining the embedding prediction module and the auxiliary speaker classifier, the final objective function of the proposed model becomes:
\begin{equation}
 \setlength{\abovedisplayskip}{3pt}
 \setlength{\belowdisplayskip}{3pt}
 L = L_{rcon} + \lambda L_{speaker\_ce} + \beta L_{embedding}
 \label{eq7}
\end{equation}
where $\lambda$ and $\beta$ are the tunable weights for the auxiliary speaker classifier loss and next embedding loss.
\vspace{-0.2cm}
\subsection{BERT encoder} 
To produce more natural prosody, we also add a BERT encoder to capture richer semantic information. Specifically, we use a pre-trained BERT~\cite{devlin2018bert} model to extract character-level embeddings to represent each character in a sentence. And the character-level BERT embedding is up-sampled to phoneme-level by simple replication, which is treated as the input of the auxiliary BERT encoder. As shown in Figure~\ref{fig:am_architecture}, the output of the auxiliary encoder is concatenated to the output of the text encoder. So the decoding process in Eq~\eqref{eq_decoder} finally becomes
\begin{equation}
 \setlength{\abovedisplayskip}{2pt}
 \setlength{\belowdisplayskip}{2pt}
 \hat{m} = d(c, s, e_{n-1}, b| \Theta{d}),
 \label{eq1}
\end{equation} where $b$ represents the output of the BERT encoder that includes rich semantic information which is helpful to the naturalness of synthesized speech~\cite{fang2019towards,hayashi2019pre}.
\vspace{-0.3cm}
\section{Spontaneous behaviors prediction and control}\label{sp}
\begin{figure}[t]
 \centering
 \includegraphics[width=0.8\linewidth]{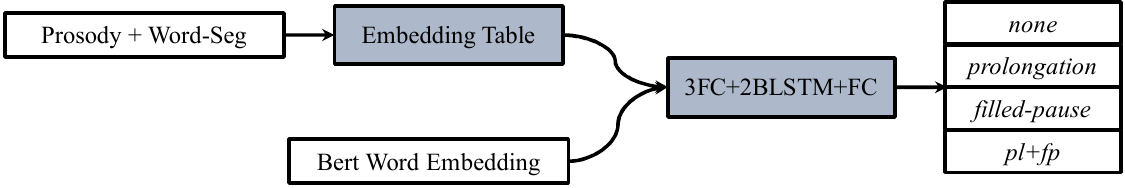}
 \caption{The architecture of spontaneous behavior predictor. The None means that no spontaneous behavior occurs at the current position. The pl+fp means that the prolongation and filled-pause occur at the same time.}
 \label{fig:spon_tag_predictor}
 \vspace{-0.2cm}
\end{figure}
The manually-labelled spontaneous behaviors are used in training, which are not readily unavailable during inference for new conversation. We need to determine the location and type of spontaneous behaviors in the text automatically. To this end, we design an individual predictor to predict the spontaneous behaviors directly from text features.

In details, we treat the prediction of behaviors as a simple classification task, where there are four available classes for each character: no spontaneous behavior after the current character (none), \textit{filled-pause}, \textit{prolongation}, or a combination of both at the same time (which sometime happens). Since the training texts with marked behavior labels are limited, we also use BERT embeddings to represent each character given texts in this predictor to alleviate the potential over-fitting problem. Besides, we also utilize the word boundary and prosody boundary information as auxiliary features to predict the spontaneous behaviors. Fig.~\ref{fig:spon_tag_predictor} shows the data flow of our spontaneous behavior predictor, where the predictor is composed of 3 fully connected layers, 2 BLSTM layers and another fully connected layer.

With a trained behavior predictor, we compute the probabilities of the four classes and then select the one with the highest probability as the predicted class. In addition, based on the predictor, to achieve controllability to the frequency of spontaneous occurring in the synthesized speech, we can use a well-designed selector to classify the text sequence. Specially, given a frequency of spontaneous behaviors and text, we can get the number $N$ that spontaneous behaviors should occur in the text. And then we use the output probabilities of the four classes and select the top $N$ spontaneous classes to label the corresponding text token. In this way, we can ensure that there are $N$ spontaneous behaviors in this sentence. We summarize the procedure in Algorithm~\ref{alg:Framwork}.
\begin{algorithm}[htb]
 \footnotesize
 \caption{Controllable spontaneous behavior selector}  
 \label{alg:Framwork} 
 \begin{algorithmic}[1]
   \Require  
     Spontaneous behavior frequency, $p$;  
     input text, $T=\{t_1, t_2, ..., t_M\}$; 
     The label $pl$ means \textit{prolongation} and $fp$ means \textit{filled-pause}.
   \Ensure  
     The spontaneous label $S=\{s_1, s_2, ... s_M\}$; 
   \State Get the number of spontaneous behaviors $N$ that should appear in this sentence. $N=int(p * M)$;  
   \State Compute the probability vector $P \in R^M$ and its corresponding labels $C \in \{pl,fp,pl+fp\}$, where $p_i = \max(p_i^{pl}, p_i^{fp}, p_i^{pl+fp})$; 
   \State Compute index set $O=\{o_1,...,o_M\}$, where $p_{o_x} > p_{o_y}$ when $x > y$;
   \For{each $i\in [0,M-1]$}
   \State \textbf{if} $(i < N):$ $s[o_i] = C_{o_i}$;
   \State \textbf{else}: $s[o_i]=none$.
   \EndFor 
   \State \Return Spontaneous behaviors label $S$;
 \end{algorithmic}  
\end{algorithm} 
\vspace{-0.7cm}
\section{Experiments}
\vspace{-0.2cm}
\subsection{Basic setups}
We conduct our experiments on a Mandarin conversation corpus in which the conversations are recorded according to the method described in \cite{guo2020conversational}. The total of 486 conversations are from two female speakers, where each conversation includes 10-20 rounds on a specific topic. It contains about 7 hour speech at 16kHz, where there are about 3,218 manually-labelled spontaneous behaviors (\textit{prolongation} and \textit{filled-pause}). We reserve 10 complete conversations, a total of 64 conversation pairs with 128 utterance as a test set. This configuration is applied to both the acoustic model and the spontaneous behavior predictor. Note that we build a two-speaker TTS model generating two-party conversations and evaluation are conducted on the synthesized conversations. Because of the relatively small set of training data, we firstly pre-train a multi-speaker acoustic model on a multi-speaker corpus to obtain a base model and then retrain this model using the conversation data. We find this procedure improves the pronunciation stability of our model.
\vspace{-0.2cm}
\subsection{Model details}
As described in Section~\ref{sec:label}, our model treats the phoneme-level hybrid linguistic features as input of the text encoder~\cite{wang2017tacotron}. For the decoder, the GMM attention~\cite{battenberg2019location} is employed to obtain stable alignments. The decoder architecture is similar to that in Tacotron2~\cite{shen2018natural}. For speaker representation, we adopt a speaker looking-up table with 256 dimension. The ``next predictor`` consists of 3 fully connected layers.
For the context acoustic encoder, we use the same configuration as described in~\cite{wang2018style}, where the number of tokens is set to 10. The speaker classifier with GRL layer contains three fully connected layers. As for the neural vocoder, we train a universal multi-band WaveRNN~\cite{kalchbrenner2018efficient,yu2020durian}. For the BERT embedding, we use a pre-trained Chinese BERT model~\cite{devlin2018bert} to extract 768-dimensional character embedding as the input of the auxiliary encoder and the spontaneous behavior predictor. 

\subsection{Evaluation on acoustic model}
Firstly, we directly utilize the conversation corpus to train a baseline acoustic model referred as M1, which does not contain the BERT-encoder and the acoustic context encoder. In the system M1, the linguistic representation does not include the explicit spontaneous behaviors label. To verify whether the spontaneous labels can clearly control the performance of spontaneous behaviors, we directly inject the spontaneous tags into the linguistic features based on M1, which is referred as M2. Besides, we further conduct experiments on the effectiveness of the BERT encoder and the acoustic context module. Finally, there are totally four models used for evaluation:
\begin{enumerate}[leftmargin=15pt,itemsep=0pt]
\item[$\bullet$] M1: baseline Taco-like model without spontaneous labels.
\item[$\bullet$] M2: baseline model with explicit spontaneous labels.
\item[$\bullet$] M3: M2 with extra BERT encoder.
\item[$\bullet$] M4: M3 with audio context encoder to model entrainment.
\end{enumerate}
We conduct CMOS test to evaluate the performance of each pair in test set of models. In order to measure the performance of conversation, we organize our test set with paired utterance like $(A_{n-1}, B_{n})$ in one conversation. There are a group of 20 native speakers taking part in each CMOS test, and they are asked to compare the naturalness and especially the entrainment of each paired utterance. The CMOS results in terms of various metrics are shown in Table~\ref{table:t1}. 

Firstly we focus on the explicit modeling with spontaneous behaviors in the evaluation ``M1 vs M2''. The result indicates that explicit labeling benefits a lot to the spontaneous conversational speech synthesis, which is intuitive since we directly learn the behaviors in a supervised way. The evaluation ``M2 vs M3'' tends to confirm the influence of the auxiliary BERT embedding. The result shows that system M3 with BERT encoder outperforms the one without (M2), which means the latent semantic information learned from a pre-trained BERT embedding can improve the performance of conversational speech synthesis. Natural conversation involves interactions of two parties, so we further evaluate the performance of chat history/context modeling by the acoustic context encoder in ``M3 vs M4''. In this evaluation, we find the performance of system M4 apparently outperforms M3, which shows the effectiveness of modeling the acoustic conversation history. This may also indicate that modeling entrainment at acoustic level between the two speakers is beneficial to the naturalness of the synthesized conversation.
\begin{table}[!htb]
 \centering
 \setlength{\belowcaptionskip}{-4.5pt}
 \caption[]{The results of CMOS evaluation. The CMOS test such as ``A vs. B'' tends to describe the degree of preference between A and B, where B is better when the CMOS score is positive and a larger positive value indicate B is much better than A. }
 \resizebox{0.75\linewidth}{!}{
   \begin{tabular}{c|cccc}
   \hline \hline
    &  & \multicolumn{3}{c}{Preference(\%)} \\ \cline{3-5} 
    & CMOS & Left & Neutral & Right \\ \hline
   M1 vs. M2 & 0.58 & 21.7 & 17 & 61.3 \\ 
   M2 vs. M3 & 0.17 & 19.8 & 42 & 38.2 \\ 
   M3 vs. M4 & 0.21 & 26.6 & 33.3 & 40.1 \\  \hline \hline
   \end{tabular}
 }
 \label{table:t1}
\end{table}
\vspace{-0.5cm}
\subsection{Evaluation on spontaneous behavior predictor}
\begin{table}[!htb]
  \centering
  \setlength{\belowcaptionskip}{-3pt}
  \caption[]{The Objective indicators: precision, recall and F-score of spontaneous predictor}
  \resizebox{0.75\linewidth}{!}{
      \begin{tabular}{cccc}
        Class & Precision & Recall & F1-score \\ \hline \hline
        \textit{none} & 0.97 & 0.98 & 0.98 \\
        \textit{filled-pause} & 0.5 & 0.48 & 0.49 \\
        \textit{prolongation} & 0.29 & 0.2 & 0.24 \\ 
        \textit{pl + fp} & 0.57 & 0.46 & 0.51 \\ \hline
      \end{tabular}
  }
  \label{table:t2}
 \end{table}

\begin{figure}[htbp]
  \centering
  \includegraphics[width=0.75\linewidth]{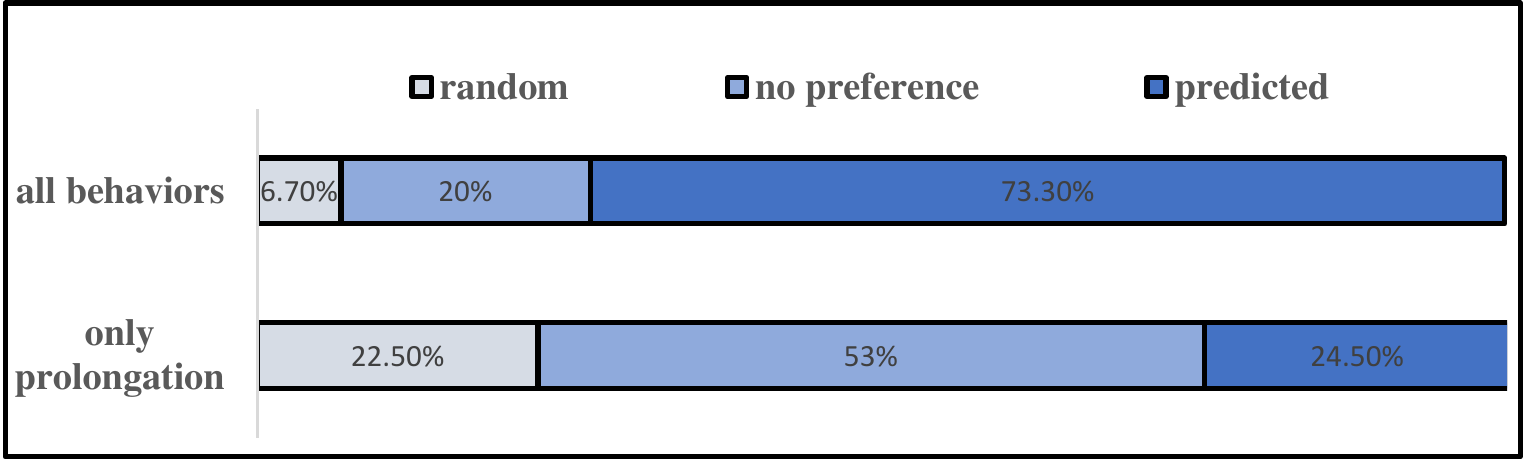}
  \caption{The AB test results of synthesized speech using random inserting and predicted inserting.}
  \label{fig:ab}
  \vspace{-3pt}
 \end{figure}

To obtain spontaneous labels during inference, we build a behavior predictor according to Section~\ref{sp}. Both objective and subjective evaluations are utilized to measure the performance of the prediction model. The objective results in Table~\ref{table:t2} show that the precision of the \textit{none} category is the highest, since most cases are \textit{none} in the training set as well as in the real conversations. As for the spontaneous behaviors, we can find that the precision is low especially for the \textit{prolongation}. But we believe this phenomenon is reasonable, since it's hard to predict spontaneous events from text~\cite{dall2014investigating}. So the goal of predicting  behaviors is to find some reasonable spontaneous positions which make the generated speech more natural. To confirm this assumption, we conduct an AB test between synthesized speech with predicted spontaneous tags or randomly inserted tags, as shown in Fig.~\ref{fig:ab}. The result show that the generated speech with predicted spontaneous labels significantly outperforms the speech with randomly arranged behaviors. As for the \textit{prolongation}, we find that it may occur at any position in the character sequence, so its precision is only 0.29. We believe the position of inserting \textit{prolongation} may not affect the naturalness of synthesis speech. We further conduct an AB test of inserting only \textit{prolongation} with predicted and randomly arranged label. Results show that the listener cannot tell the obvious difference in the naturalness of both, which verifies our assumption.

\begin{figure}[ht]
  \centering
  \includegraphics[width=0.75\linewidth]{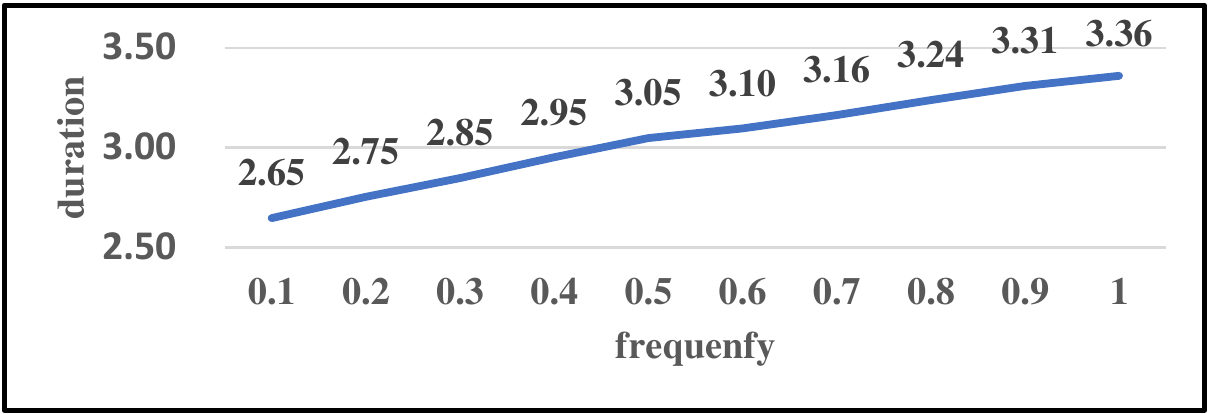}
  \caption{The impact of different spontaneous behaviour occurrence frequencies on sentence average duration.}
  \label{fig:fq}
  \vspace{-3pt}
 \end{figure}

For the control to the spontaneous behaviors, we synthesize speech with different spontaneous behavior frequency from 0.1 to 1.0 based on the selection Algorithm~\ref{alg:Framwork}. We count the average duration of synthesized speech corresponding to different spontaneous frequencies and the result is shown in Fig.\ref{fig:fq}. We can see that as the frequency increases, the corresponding duration increases and the synthesized audio becomes less fluent\footnote{We suggest the readers to listen to our samples at \url{https://syang1993.github.io/spon_tts}}.
\vspace{-0.2cm}
\vspace{-0.8cm}
\vspace{-1cm}
\section{Conclusions}
This paper proposes a unified controllable spontaneous conversational speech synthesis framework, which directly models two typical spontaneous behaviors \textit{prolongation} and \textit{filled-pause}. In the proposed model, we utilize an acoustic context encoder to model the entrainment in conversation and a BERT encoder to extract semantic information from texts, which significantly improve the performance of  conversational speech synthesis. To achieve flexible control during inference, we also propose a behavior predictor to determine the positions and types of spontaneous behaviors from, which can be used to control the degree of fluency in the synthesized speech. 

\bibliographystyle{IEEEtran}
\bibliography{template}
\end{CJK*}
\end{document}